**Topical Issue in *Solar Physics*: Flux-rope Structure of Coronal Mass Ejections**

**Preface**

N. Gopalswamy[1], T. Nieves-Chinchilla[1,2], M. Hidalgo[3], J. Zhang[4], P. Riley[5], and L. van Driel-Gesztelyi[6,7,8], C.H. Mandrini[9]

[1]Heliophysics Division, NASA Goddard Space Flight Center, Greenbelt, MD, USA
[2]Department of Physics, The Catholic University of America, Washington DC 20064, USA
[3]SRG-UAH, Alcala de Henares, Madrid, Spain
[4]School of Physics, Astronomy and Computational Sciences, George Mason University, Fairfax, VA, USA
[5] Predictive Science Inc., San Diego, CA, USA
[6]Observatoire de Paris, LESIA-CNRS UMR 8109, 9295 Meudon Cédex, France
[7]University College London, Mullard Space Science Laboratory, Holmbury St. Mary, UK
[8]Konkoly Observatory of Hungarian Academy of Sciences, Budapest, Hungary
[9]Instituto de Astronomía y Física del Espacio, CONICET-UBA, CC. 67, Suc. 28, 1428 Buenos Aires, Argentina

This Topical Issue of *Solar Physics*, devoted to the study of flux-rope structure in coronal mass ejections (CMEs), is based on two Coordinated Data Analysis Workshops (CDAWs) held in 2010 (20 - 23 September in Dan Diego, California, USA) and 2011 (September 5-9 in Alcala, Spain). The primary purpose of the CDAWs was to address the question: Do all CMEs have flux rope structure? Each CDAW was attended by about 50 scientists interested in the origin, propagation and interplanetary manifestation of CME phenomena.

The backbone of the workshop was a set of 59 interplanetary CMEs (ICMEs) that were driving shocks at Sun-Earth L1 as detected by one or more of the *Solar Heliospheric Observatory* (SOHO), *Wind*, and the *Advanced Composition Explorer* (ACE). The CME-ICME pairs were selected from a set identified by Gopalswamy *et al*. (2010) based on the criterion that the source location should be within $\pm 15^\circ$ longitude from the disk center. Many of the papers in this topical issue (TI) utilized the CME-ICME pairs, referred to as the CDAW Events. A revision of the source locations made during the CDAWs reduced the list to 54 events. According to the classical definition of Burlaga *et al*. (1981), 23 ICMEs were classified as magnetic clouds (MCs) and the remaining 31 were non-MCs. The reason for limiting the longitude range to $\pm 15^\circ$ is that disk-center CMEs are more likely to be identified as MCs according to the geometrical hypothesis: all ICMEs are flux ropes but appear as non-ropes due to observational limitations. The events all occurred during Solar Cycle 23 (bounded by the launch of SOHO to the end of 2005) with exceptional *in-situ* measurements and remote-sensing observations for each. The remote-sensing observations include H$\alpha$, EUV, and X-ray images from ground and space based instruments.

Yashiro *et al*. (2013) focused on the structure of post-eruption arcades (PEAs) associated with MC and non-MC CMEs and found that one cannot distinguish between these two classes of events based on flare data. Gopalswamy *et al*. (2013) compared the Fe and O charge states in MCs and non-MCs and found that enhanced charge state is a common characteristic of both types of ICMEs. They also concluded that the non-rope models involving magnetic loop expansion are inconsistent with non-MCs because the observed charge state and CME kinematics do not support such a model. Xie, Gopalswamy, and St. Cyr (2013) were able to fit a flux rope to CMEs associated with both MCs and non-MCs and showed evidence that the propagation effects might turn them into MCs and non-MCs; specifically, that CMEs associated with non-MCs are generally deflected away from the Sun-Earth line, while those associated with MCs were unaffected or were deflected towards the Sun-Earth line (Mäkelä *et al*. 2013). This result was also supported by the fact that the direction parameter is larger for CMEs associated with MCs than for the non-MC CMEs (Kim *et al*. 2013). Zhang, Hess, and Poomvises (2013) presented on a case study of two ICMEs and also concluded that the difference between the two events observed *in-situ* can be explained by the deflection of flux ropes *en route* to Earth. Cho et al. (2013) determined the helicity signs in the source active regions of the CDAW events by estimating the cumulative magnetic helicity injected through the photosphere. They found that in 88% of the cases, the ICME helicity signs are consistent with those of the solar source regions. They also suggested that one or more of the following could have caused the deviation in the remaining cases: incorrect identification of the CME source region, local helicity sign opposite to that of the entire active region, and the helicity sign of the pre-existing coronal magnetic field opposite to the sign of the photospheric helicity injection.

All the CDAW events were analyzed using four different magnetic field models and reconstruction techniques: force-free fitting, magnetostatic reconstruction using a numerical solution to the Grad-Shafranov equation, fitting to a self-similarly expanding cylindrical configuration and elliptical, non-force-free fitting (Al-Haddad *et al*. 2013). Hidalgo, Nieves-Chinchilla, and Blanco (2013) used an analytical flux rope model to fit the observations and found that the majority of CDAW events contain flux ropes. They also found that the flux-rope noses are generally along the Sun-Earth line. Blanco *et al*. (2013) studied the Forbush decrease in cosmic rays triggered by the passage of the CDAW events at Earth and found that only 25 % displayed a noticeable decrease. They also found that MCs are more effective in causing Forbush decreases.

The TI also includes papers that expand the context of the CDAW events: Vourlidas *et al*. (2013) present a statistical analysis of all white-light CMEs observed by SOHO, assisted by 3D MHD simulations. They suggest that a flux rope can be defined as a coherent magnetic, twist-carrying coronal structure with angular width of at least 40º and able to reach beyond 10 Rs. Isavnin, Vourlidas, and Kilpua (2013) studied 15 ICMEs in Cycle 24, comparing the three-dimensional parameters of CMEs from imaging and *in-situ* reconstructions, and focusing on propagation effects. They were able to confirm the flux rope deflection towards the equator and its rotation. Riley and Richardson (2013) analyzed *Ulysses* spacecraft measurements to assess five possible explanations for why some ICMEs are observed to be MCs and others are not. They concluded that it is difficult to choose between the geometrical hypothesis discussed above and the possibility that there exists two distinct initiation mechanisms - one producing MCs and the other non-MCs. Romashets and Vandas (2013) considered a linear force-free configuration consisting of a cylindrical flux rope combined with a compact toroid. This model can be applied for the interpretation of some features observed in solar flux ropes, including prominences. Berdichevsky (2013) studied the isotropic evolution of flux ropes and attempted to estimate the mass of ICMEs. Osherovich, Fainberg, and Webb (2013) provided observational support for the presence of double helix structure within CMEs and MCs. Hu *et al*. (2013) examined the effect of electron pressure on the Grad-Shafranov reconstruction of ICMEs and found that it contributes to a 10 - 20 % discrepancy in derived physical quantities, such as the magnetic flux content of the ICME flux rope observed at 1 AU.

As in the cases of previous CDAWs, the data collected for the Flux Rope CDAWs are available on line: http://cdaw.gsfc.nasa.gov/ meetings/2010_fluxrope/LWS_CDAW2010_ICMEtbl.html. We would like to express our gratitude to the referees whose critical assessment of the manuscripts resulted in the high quality papers in this TI "Flux-rope structure of coronal mass ejections". The CDAW organizers gratefully acknowledge support for the CDAWs from NASA's Living with a Star (LWS) program. We thank Lika Guhathakuta, at NASA Head Quarters, for her assistance. We also thank Predictive Science, Inc. (USA) and the University of Alcala (Spain) for hosting the CDAWs in 2010 and 2011, respectively.

## References

Al-Haddad, N., Nieves-Chinchilla, T., Savani, N. P., Möstl, C., Marubashi, K., Hidalgo, M. A., Roussev, I. I., Poedts, S., Farrugia, C. J.: 2013, *Solar Phys*. this issue. doi: 10.1007/s11207-013-0244-5.
Berdichevsky, D.: 2013, *Solar Phys*. this issue. doi: 10.1007/s11207-012-0176-5.
Blanco, J. J., Catalán, E., Hidalgo, M. A., Medina, J., García, O., Rodríguez-Pacheco, J.: 2013, *Solar Phys*. this issue. doi: 10.1007/s11207-013-0256-1.
Burlaga, L., Sittler, E., Mariani, F., Schwenn, R.: 1981, *J. Geophys. Res*. 86, 6673, doi: 10.1029/JA086iA08p06673.
Cho, K.-S., Park, S.-H., Marubashi, K., Gopalswamy, N., Akiyama, S., Yashiro, S., Kim, R.-S., Lim, E.-K.: 2013, *Solar Phys*. this issue. doi: 10.1007/s11207-013-0224-9.
Gopalswamy, N., Xie, H., Mäkelä, P., Akiyama, S., Yashiro, S., Kaiser, M. L., Howard, R. A., Bougeret, J.-L.: 2010, *Astrophys. J.* 710, 1111, doi: 10.1088/0004-637X/710/2/1111.
Gopalswamy, N., Mäkelä, P., Akiyama, S., Xie, H., Yashiro, S., Reinard, A. A.: 2013, *Solar Phys*. this issue. doi: 10.1007/s11207-012-0215-2.
Hidalgo, M. A., Nieves-Chinchilla, T., Blanco, J. J.: 2013, *Solar Phys*. this issue. doi: 10.1007/s11207-012-0191-6.
Hu Q., Farrugia, C.J., Osherovich, V.A., Möstl, C., Szabo, A., Oglivie, K.W., Lepping, R.P.: 2013, *Solar Phys.* this issue. doi: 10.1007/s11207-013-0259-y.
Isavnin, A., Vourlidas, A., Kilpua, E. K. J.: 2013, *Solar Phys*. this issue. doi: 10.1007/s11207-012-0214-3.
Kim, R.-S., Gopalswamy, N., Cho, K.-S., Moon, Y.-J., Yashiro, S.: 2013, *Solar Phys*. this issue. doi: 10.1007/s11207-013-0230-y
Mäkelä, P., Gopalswamy, N., Xie, H., Mohamed, A. A., Akiyama, S., Yashiro, S.: 2013, *Solar Phys*. this issue. doi: 10.1007/s11207-012-0211-6.
Osherovich, V., Fainberg, J., Webb, A.: 2013, *Solar Phys.* this issue. doi: ...
Riley, P. and Richardson, I. G.: 2013, *Solar Phys*. this issue. doi: 10.1007/s11207-012-0006-9,
Romashets, E., Vandas, M.: 2013, *Solar Phys*. this issue. doi: 10.1007/s11207-012-0083-9.
Vourlidas, A., Lynch, B.J., Howard, R.A., Li, Y.: 2013, this issue, doi: 10.1007/s11207-012-0084-8.

Xie, H., Gopalswamy, N., St. Cyr, O. C.: 2013, *Solar Phys*. this issue. doi: 10.1007/s11207-012-0209-0.
Yashiro, S., Gopalswamy, N., Mäkelä, P., Akiyama, S.: 2013, Solar Phys. this issue. doi: 10.1007/s11207-013-0248-1.
Zhang, J., Hess, P., Poomvises, W.: 2013, *Solar Phys*. this issue. doi: 10.1007/s11207-013-0242-7.